\RequirePackage{fix-cm}
\documentclass[twocolumn,epjc3]{svjour3}  
\usepackage{hyperref}
\usepackage{cite}
\usepackage{amsmath}
\usepackage{amssymb}
\usepackage{amsfonts}
\usepackage{appendix}
\usepackage{orcidlink}

\usepackage{lipsum}     
\usepackage{mathtools}
\usepackage{cuted}

\usepackage{graphics}
\usepackage{latexsym}
\usepackage{orcidlink}

\smartqed  
\RequirePackage{graphicx}
\journalname{Chinese Journal of Physics}
\def\ref@jnl#1{{\jnl@style#1}}%
\newcommand\aj{\ref@jnl{AJ}}
\newcommand\psj{\ref@jnl{PSJ}}
\newcommand\araa{\ref@jnl{ARA\&A}}
\newcommand\apjl{\ref@jnl{ApJL}}     
\newcommand\apjs{\ref@jnl{ApJS}}
\newcommand\apss{\ref@jnl{Ap\&SS}}
\newcommand\aap{\ref@jnl{A\&A}}
\newcommand\aapr{\ref@jnl{A\&A~Rv}}
\newcommand\aaps{\ref@jnl{A\&AS}}
\newcommand\azh{\ref@jnl{AZh}}
\newcommand\baas{\ref@jnl{BAAS}}
\newcommand\icarus{\ref@jnl{Icarus}}
\newcommand\jaavso{\ref@jnl{JAAVSO}}  
\newcommand\jrasc{\ref@jnl{JRASC}}
\newcommand\memras{\ref@jnl{MmRAS}}
\newcommand\mnras{\ref@jnl{MNRAS}}
\newcommand\prd{\ref@jnl{PhRvD}}
\newcommand\prl{\ref@jnl{PhRvL}}
\newcommand\pasp{\ref@jnl{PASP}}
\newcommand\pasj{\ref@jnl{PASJ}}
\newcommand\qjras{\ref@jnl{QJRAS}}
\newcommand\skytel{\ref@jnl{S\&T}}
\newcommand\solphys{\ref@jnl{SoPh}}
\newcommand\sovast{\ref@jnl{Soviet~Ast.}}
\newcommand\ssr{\ref@jnl{SSRv}}
\newcommand\zap{\ref@jnl{ZA}}
\newcommand\iaucirc{\ref@jnl{IAUC}}
\newcommand\aplett{\ref@jnl{Astrophys.~Lett.}}
\newcommand\apspr{\ref@jnl{Astrophys.~Space~Phys.~Res.}}
\newcommand\bain{\ref@jnl{BAN}}
\newcommand\fcp{\ref@jnl{FCPh}}
\newcommand\gca{\ref@jnl{GeoCoA}}
\newcommand\grl{\ref@jnl{Geophys.~Res.~Lett.}}
\newcommand\jgr{\ref@jnl{J.~Geophys.~Res.}}
\newcommand\jqsrt{\ref@jnl{JQSRT}}
\newcommand\memsai{\ref@jnl{MmSAI}}
\newcommand\nphysa{\ref@jnl{NuPhA}}
\newcommand\physrep{\ref@jnl{PhR}}
\newcommand\physscr{\ref@jnl{PhyS}}
\newcommand\planss{\ref@jnl{Planet.~Space~Sci.}}
\newcommand\procspie{\ref@jnl{Proc.~SPIE}}

\newcommand\actaa{\ref@jnl{AcA}}
\newcommand\caa{\ref@jnl{ChA\&A}}
\newcommand\cjaa{\ref@jnl{ChJA\&A}}
\newcommand\jcap{\ref@jnl{JCAP}}
\newcommand\na{\ref@jnl{NewA}}
\newcommand\nar{\ref@jnl{NewAR}}
\newcommand\pasa{\ref@jnl{PASA}}
\newcommand\rmxaa{\ref@jnl{RMxAA}}

\newcommand\maps{\ref@jnl{M\&PS}}
\newcommand\aas{\ref@jnl{AAS Meeting Abstracts}}
\newcommand\dps{\ref@jnl{AAS/DPS Meeting Abstracts}}

\begin{document}

\title{Tidal Love numbers of anisotropic stars within the complexity factor formalism
}


\author{
        \'Angel Rinc{\'o}n \orcidlink{0000-0001-8069-9162} \thanksref{e1,addr1}
        \and
        Grigoris Panotopoulos\orcidlink{0000-0003-1449-9108}
        \thanksref{e2,addr2}
        \and
        Ilidio Lopes \orcidlink{0000-0002-5011-9195} \thanksref{e3,addr3}
        \and
}

\thankstext{e1}{e-mail: 
\href{mailto:angel.rincon@ua.es}
{\nolinkurl{angel.rincon@ua.es}}
}

\thankstext{e2}{e-mail: 
\href{mailto:grigorios.panotopoulos@ufrontera.cl}
{\nolinkurl{grigorios.panotopoulos@ufrontera.cl}}
}

\thankstext{e3}{e-mail: 
\href{mailto:ilidio.lopes@tecnico.ulisboa.pt}
{\nolinkurl{ilidio.lopes@tecnico.ulisboa.pt}}
}


\institute{Departamento de Física Aplicada, Universidad de Alicante,
           Campus de San Vicente del Raspeig, E-03690 Alicante, Spain. \label{addr1}
           \and
           Departamento de Ciencias F{\'i}sicas, Universidad de la Frontera, Casilla 54-D, 4811186 Temuco, Chile.
           \label{addr2}
           \and
           Centro de Astrof{\'i}sica e Gravita{\c c}{\~a}o, Departamento de F{\'i}sica, Instituto Superior T{\'e}cnico-IST, Universidade de Lisboa-UL, Av. Rovisco Pais, 1049-001 Lisboa, Portugal. 
           \label{addr3}
}

\date{Received: date / Accepted: date}

\maketitle

\begin{abstract}
We compute the quadrupolar gravitoelectric tidal Love numbers of spherical configurations made of anisotropic matter. Anisotropies are introduced within the vanishing complexity factor, while interior solutions are obtained adopting the Extended Chaplygin gas equation-of-state. A comparison with a more conventional approach is made as well.
\keywords{
Relativistic stars \and Complexity factor \and Anisotropic stars}
\end{abstract}

\section{Introduction}\label{intro}

\smallskip\noindent

The prevailing questions in modern fundamental physics, concerning the possible existence of unknown particles or even unknown properties related to gravitation, compel us to seek new methods for exploring these mysteries. In this study, we adopt a novel approach – employing the complexity formalism – to discover anisotropic solutions to Einstein's Field Equations within a stellar context, concurrently exploring new classes of particles and stellar objects. We begin this introduction by highlighting some important results pertinent to our study:
\begin{itemize}
\item[$-$] Dark matter, constituting approximately 27\% of the universe's total mass-energy content, remains an enigmatic subject in fundamental physics, encompassing particle physics, astrophysics, and cosmology \cite{2017arXiv170704591B,2022RPPh...85e6201B,2014IJMPD..2330005D}. Its presence, inferred from astrophysical and cosmological observations, manifests through gravitational effects on galaxies and cosmic structures, including the cosmic microwave background radiation \cite{2017FrPhy..12l1201Y}. Prevailing theories propose that dark matter's components exist outside the Standard Model of particle physics, suggesting the existence of yet-undiscovered particles. This spectrum includes a range of hypothetical candidates such as Fuzzy Dark Matter, axion-like particles, Axions, Sterile Neutrinos, and Weakly Interacting Massive Particles \cite{2010ARA&A..48..495F,2015PhR...555....1B,2018RPPh...81f6201R,2019Univ....5..213P}.
	
\smallskip\noindent

\item[$-$] Compact objects, resulting from stellar evolution, represent a significant deviation from conventional matter. Their immense internal densities demand exploration through Einstein’s General Relativity (GR) \cite{Einstein:1915ca,Shapiro:1983du,Psaltis:2008bb,Lorimer:2008se,Zorotovic:2019uzl}. Neutron stars, in particular, require a multidisciplinary approach involving nuclear particle physics, astrophysics, and gravitational physics. As the densest celestial bodies, second only to black holes, they mimic conditions unattainable in Earth-based laboratories, serving as unique cosmic laboratories. Within this category, strange quark stars — currently theoretical — could represent the fundamental state of hadronic matter \cite{Alcock:1986hz,Alcock:1988re,Madsen:1998uh,Weber:2004kj,Yue:2006it,Leahy:2007we}.
	
\smallskip\noindent

\item[$-$] Recent research has speculated on the presence of strange matter in the cores of neutron star-hybrid stars \cite{Benic:2014jia,Yazdizadeh_2022,EslamPanah:2018rfe}, with some studies suggesting similarities to conventional neutron stars \cite{Jaikumar:2005ne}. Strange quark stars might also play a key role in explaining the phenomenon of super-luminous supernovae \cite{Ofek:2006vt,2009arXiv0911.5424O}. Their potential existence across various astrophysical settings has spurred recent research \cite{Mukhopadhyay:2015xhs,Panotopoulos:2018ipq}.
\end{itemize}

\smallskip\noindent

In keeping with our exploratory approach, we have chosen to study dark energy stars — a class of objects currently under investigation that could offer invaluable insights into the roles of dark matter and dark energy in astrophysics and cosmology. Indeed, dark energy stars \cite{Rahaman:2011hd} have been discussed in various scenarios, among which the following are particularly noteworthy:
these stars offer a possible alternative explanation for observations currently attributed to compact stars including  black holes\cite{10.1007/s10714-021-02851-x};  dark energy stars  also might affect the characteristics of neutron stars \cite{10.1088/1475-7516/2019/07/012}, helping us detect dark energy and better understand its properties\cite{10.1088/1475-7516/2021/02/045,10.1088/1475-7516/2012/03/037}; the existence of  dark energy stars is relevant for understanding the distribution of dark energy in the vicinity of compact objects, such as stars and black holes, and its potential impact on their dynamics \cite{10.48550/arxiv.1412.7323}. Indeed, dark energy stars  could reshape our understanding of black holes, the distribution of dark energy, and the fundamental forces that shape the universe.

In particular, it is worth noting that almost two decades ago the concept of dark energy stars emerged, positing the gravitational collapse of entities larger than a few solar masses, culminating in the formation of a compact entity. These objects have a surface that coincides with a quantum-critical surface for space-time. Inside, their interior deviates from conventional space-time merely by containing significantly increased vacuum energy \cite{Chapline:2004jfp}.
Dark energy stars could explain where high-energy cosmic rays and positrons come from. As masses fall into them, they break down into lighter particles near the edge, accelerating the decay of protons. This idea could change our current understanding of space events such as supernova explosions, gamma-ray bursts, the release of positrons, and even the mysteries of dark matter.

\smallskip\noindent

This research delves into the hypothesized connection between dark energy and compact stars, exploring this notion through two novel concepts: the potential anisotropy of dark energy and the vanishing complexity factor formalism within General Relativity \cite{Einstein:1915ca}. This approach addresses limitations in previous models, such as the oversight of certain elements in the energy density fluid and the substitution of probability distributions with the fluid's energy density \cite{Sanudo:2008bu,Herrera:2018bww}.

\smallskip\noindent

The isotropic fluid, typically used to model compact stars, offers uniform radial pressure. However, this is not always the case. Anisotropy, where pressure varies directionally, is often found. Ruderman's seminal work \cite{Ruderman:1972aj} proposed that specific nuclear matter interactions could cause this uneven pressure. Subsequent studies, such as \cite{Bowers:1974tgi}, extended this idea, showing that conditions like phase transitions or the presence of certain super-fluids could lead to anisotropies.

\smallskip\noindent

Recent advancements in this field have been marked by significant contributions, including the exploration of anisotropic quark stars. Pioneering studies like \cite{MakANDHarko} have provided exact analytical solutions, while \cite{Deb:2016lvi} and \cite{Deb:2015vda} have furthered our understanding through non-singular solutions and the Homotopy Perturbation Method. Efforts to merge anisotropic features into existing isotropic models have also been notable, as seen in \cite{Gabbanelli:2018bhs,Ovalle:2017fgl,Ovalle:2017wqi}.

\smallskip\noindent

Anisotropic matter plays a unique role in the cosmos, influencing the behavior and structure of astronomical objects like neutron stars. This characteristic is crucial for understanding phenomena such as dark matter. For instance, dark matter clouds might exhibit local anisotropy, akin to collisionless particle systems. This topic is comprehensively covered in \cite{2022NewAR..9501662K}. In General Relativity, anisotropic fluids are well-studied, with their ability to follow geodesic paths as highlighted by \cite{2002JMP....43.4889H}. Significant research has also been conducted on their impact on the mass-radius relationship of anisotropic stars \cite{2003RSPSA.459..393M}.

 \smallskip\noindent
 
Our investigation hinges on two main concepts related with  anisotropy:
(i) In the context of TOV equations, matter contributes a distinct radial pressure given by \(P_{r}(r) = P (r)\).
(ii) The existence of an anisotropic energy-momentum tensor gives rise to a specific tangential pressure: \(P_{\perp}(r)\).
We further elucidate the rationale behind these terms, representing the matter distribution using the expression \(\Pi (r)=P_{\perp}(r) - P_{r}(r)\). Our goal is to discern interior solutions for relativistic stars, focusing on quark stars composed of anisotropic matter through the lens of the complexity factor formalism. Initially examined from a purely mathematical perspective, the actual value of the complexity factor emerges when it serves as an auxiliary condition in the differential equations of self-gravitating systems. It offers a systematic approach to integrating anisotropies, particularly in ultra-dense environments found in relativistic objects \cite{Sharif:2018pgq,Sharif:2018efi,Abbas:2018cha,Herrera:2019cbx}.

\smallskip\noindent	

The inspirals and relativistic collisions between the primary and the secondary object in a binary system, and the gravitational wave (GW) signal emitted during those processes, contain a wealth of information on the inner structure and equation-of state of the colliding bodies. The imprint of the equation-of-state within the signals emitted during binary coalescences is mainly determined by adiabatic tidal fenomena that are characterized by a set of coefficients, known as the tidal deformabilities and the corresponding tidal Love numbers \cite{Love1, Love2}.

\smallskip\noindent	

Furthermore, leveraging current and upcoming gravitational wave data, we predict tidal Love numbers for this class of models, as they provide insights into gravitational interactions within a two-body system. Notably, the gravitational potential of the secondary body incorporates the conventional Newtonian potential and corrections from the radius-to-distance ratio \cite{1933MNRAS..93..449C}.
In-depth treatments of tidal Love numbers can be found in \cite{Hinderer:2007mb,Damour:2009vw}. The tidal Love number \(k\) is inherently associated with two deformability indices $\tilde{\lambda}_{1,2}$, as further elaborated in \cite{Postnikov:2010yn}. Analyzing $\tilde{\lambda}_{1,2}$ proves instrumental in investigating post-merger GW signals \cite{Flanagan:2007ix}. GW170817 data provides compelling evidence of $\tilde{\lambda}_{1,2}$'s influence, modulating the GW waveform \cite{2017PhRvL.119p1101A}. Thus, for binary systems, pinpointing $\tilde{\lambda}_{1,2}$ has cascading implications on multiple fronts, making it a central pivot for our research's forthcoming aspects.

\smallskip\noindent	

Since the tidal Love number measures how a star responds to tidal forces, it can be utilized to distinguish dark energy stars from other compact stars due to its sensitivity to the compactness parameter $M/R$ and the internal structure of the star \cite{10.1103/physrevd.82.024016,10.1103/physrevd.105.043013}.	 This makes it a powerful observational tool for identifying dark energy stars, as their unique internal structure, leads to a different Love number compared to standard  compact stars, such as neutron or quark stars.

\smallskip\noindent

The outline of our work is the following:
After this compact introduction, in the next section, we review the description of compact stars in the framework of Einstein's theory, while in section 3 we summarize the main ingredient behind the vanishing complexity formalism. Next, in the fourth section, we show the ingredients needed to understand the tidal love numbers. Subsequently, in the fifth section, we discuss our numerical results, and finally, we finish our work with some concluding remarks in section 6.  Along with this manuscript we have used geometrical units where $G=1=c$, and we adopt the mostly negative metric signature in four space-time dimensions, $\{+,-,-,-\}$.

\section{Equations for relativistic stars in General Relativity}\label{GR}

This section is devoted to summarizing the minimum ingredients useful for describing anisotropic stars in four dimensions.
Consider a static and spherically symmetric object bounded by a spherical surface $\Sigma$. Let us also assume that the object is anisotropic in the sense that there is not only radial pressure, $P_r$, but also tangential pressure, $P_{\perp}$.
It should be noted that in some cases the anisotropic nature of certain sources must be taken into account by including an additional parameter: viscosity. However, for the sake of simplicity, our analysis will be restricted to an anisotropic fluid, albeit without the contribution of viscosity, i.e. $T^{\mu}_{\nu} \equiv \text{diag} \{ \rho(r), P_{r}(r), P_{\perp}(r), P_{\perp}(r) \}$.
In what follows, we will assume a line element written in Schwarzschild--like coordinates, i.e., 
\begin{equation}
\mathrm{d}s^2 = e^{\nu} \mathrm{d}t^2 - e^{\lambda} \mathrm{d}r^2 -
r^2 \mathrm{d}\Omega^2.
\label{metric}
\end{equation}
The metric potentials, defined as $\nu(r)$ and $\lambda(r)$, depend on the radial coordinate only, and $d\Omega^2\equiv \left( d\theta^2 + \sin^2\theta d\phi^2 \right)$ represent the element of solid angle.
We will take: $x^0=t; \, x^1=r; \, x^2=\theta; \, x^3=\phi$.
The Einstein field equations, for a vanishing cosmological constant, take the simple form:
\begin{equation}
G^{\mu}_{\nu} = - 8\pi T^{\mu}_{\nu},
\label{Efeq}
\end{equation}
In the co-moving frame, the components of the energy--momentum tensor in (local) Minkowski coordinates and in terms of the metric potentials, is written as follows
\begin{eqnarray}
\rho &=& -\frac{1}{8\pi}\left[-\frac{1}{r^2}+e^{-\lambda}
\left(\frac{1}{r^2}-\frac{\lambda'}{r} \right)\right],
\label{fieq00}
\\
P_r &=& -\frac{1}{8\pi}\left[\frac{1}{r^2} - e^{-\lambda}
\left(\frac{1}{r^2}+\frac{\nu'}{r}\right)\right],
\label{fieq11}
\\
P_\bot &=& \frac{1}{32\pi}e^{-\lambda}
\left(2\nu''+\nu'^2 -
\lambda'\nu' + 2\frac{\nu' - \lambda'}{r}\right),
\label{fieq2233}
\end{eqnarray}
here primes symbols represent the derivatives with respect to $r$. 
As occurs in isotropic cases, we can obtain the anisotropic version of the hydrostatic equilibrium equation, usually known as the generalized Tolman-Opphenheimer-Volkoff (TOV) equation. To find the concrete differential equation, we 
combine Eqs.\ref{fieq00}, \ref{fieq11} and \ref{fieq2233} to get: 
\begin{equation} \label{Prp}
P'_r = -\frac{1}{2}\nu'\left( \rho + P_r\right) + \frac{2}{r}\left(P_\bot-P_r\right).  
\end{equation}
The $\nu'$-dependence in equation (\ref{Prp}) can be removed by taking advantage of the following relation
\begin{equation}
\frac{1}{2}\nu' =  \frac{m + 4 \pi P_r r^3}{r \left(r - 2m\right)},
\label{nuprii}
\end{equation}
and we then substitute the \eqref{nuprii} into \eqref{Prp} to obtain the generalized TOV, namely
\begin{equation} \label{ntov}
P'_r=-\frac{(m + 4 \pi P_r r^3)}{r \left(r - 2m\right)}\left( \rho + P_r\right)+\frac{2}{r}\left(P_\bot-P_r\right).
\end{equation}
At this point, we should highlight that the last equation can be re-interpreted as the balance between the following three forces: 
  i) gravitational, $F_g$, 
 ii) hydrostatic, $F_r$, and 
iii) anisotropic, $F_p$, 
defined according to the expressions
\begin{subequations}
\begin{align}
F_g &= -\frac{1}{2}\nu'\left( \rho+ P_r\right) ,
\\
F_r &= -P'_r ,
\\
F_p &= \frac{2\Pi}{r} .
\end{align}
\end{subequations}
The anisotropic factor, $\Delta$, is defined as usual, i.e.,  $\Delta \equiv \Pi=P_\bot-P_r$.
Utilizing the last definitions, the equation \eqref{ntov}, now can be written as 
\begin{equation}
F_g + F_r + F_p = 0,\label{Frp}
\end{equation}	
as is mentioned in Ref. \cite{Prasad:2021eju}.  
Also, notice that if $F_p=0$, we obtain the standard TOV equation, as it should be.
On one hand, when $P_\bot>P_r$  (or $\Pi>0$),  $F_p>0$ produces a repulsive force in equation (\ref{Frp}) that counteracts the attractive force given by the combination $F_g+F_r$.
On the other hand, when $P_\bot<P_r$ (or $\Pi<0$),  $F_p<0$ is also an attractive force that contributes to the other ones.

\noindent Notice that $m$ is the  mass function, obtained by:
\begin{equation}
R^3_{232}=1-e^{-\lambda}=\frac{2m}{r},
\label{rieman}
\end{equation}
or, in terms of the density
\begin{equation}
m = 4\pi \int^{r}_{0} \tilde r^2\rho \  d\tilde r.
\label{m}
\end{equation}
The energy-momentum tensor can be rewritten as follows
\begin{equation}
T^{\mu}_{\nu}=\rho u^{\mu}u_{\nu}-  P
h^{\mu}_{\nu}+\Pi ^{\mu}_{\nu},
\label{24'}
\end{equation}
where we need to introduce some additional functions. First, we introduce the  four-velocity and four acceleration as 
\begin{subequations}
\begin{align}
u^{\mu} &= (e^{-\frac{\nu}{2}},0,0,0),
\\
a^\alpha &= u^\alpha_{;\beta}u^\beta,
\end{align}
\end{subequations}
where the non--vanishing component is $a_1 = -\nu^{\prime}/2$. In addition, 
the set $\{ \Pi^{\mu}_{\nu}, \Pi, h^{\mu}_{\nu}, s^{\mu}, P \}$ is defined as
\begin{subequations}
\begin{eqnarray}
\Pi^{\mu}_{\nu} &=& \Pi\bigg(s^{\mu}s_{\nu}+\frac{1}{3}h^{\mu}_{\nu}\bigg) ,
\\
\Pi &=& P_{\bot}-P_r , \label{Delta}
\\
h^\mu_\nu &=& \delta^\mu_\nu-u^\mu u_\nu ,
\\
s^{\mu} &=& (0,e^{-\frac{\lambda}{2}},0,0),  \label{ese}
\\
P & \equiv & \frac{1}{3}\Bigl( P_{r}+2P_{\bot} \Bigl), 
\end{eqnarray}
\end{subequations}
with the properties
$s^{\mu}u_{\mu}=0$,
$s^{\mu}s_{\mu}=-1$.
We match the problem with Schwarzschild spacetime, i.e.,
\begin{equation}
\mathrm{d}s^2= \left(1-\frac{2M}{r}\right) \mathrm{d}t^2 - \left(1-\frac{2M}{r}\right)^{-1} \mathrm{d}r^2 -
r^2  \mathrm{d}\Omega^2.
\label{Vaidya}
\end{equation}
Finally, the problem is closed by imposing appropriated  boundary conditions on the surface
$r=r_\Sigma \equiv  R$. 
Taking into account the first and the second fundamental
forms across that surface we have
\begin{subequations}
\begin{eqnarray}
e^{\nu(r)} \Bigl|_{r=R} &=& 1-\frac{2M}{R},
\label{enusigma}
\\
e^{\lambda(r)} \Bigl|_{r=R} &=& \Bigg(1-\frac{2M}{R}\Bigg)^{-1},
\label{elambdasigma}
\\
P_r(r) \Bigl|_{r=R} &=& 0.
\label{PQ}
\end{eqnarray}
\end{subequations}
The last three equations are the necessary and sufficient conditions for a smooth matching of the two metrics (\ref{metric}) and (\ref{Vaidya}) on the surface $r=R$.

\section{Vanishing complexity factor formalism}\label{VCF}

This section offers readers the essential components necessary to comprehend the concept of the vanishing complexity factor formalism and its potential applications within the context of relativistic stars and astrophysical scenarios.
It is widely acknowledged that anisotropic interior solutions always necessitate an additional condition to fully describe the system.
In this regard, while there are several possible approaches to fulfill this requirement, we will adopt the concept of complexity as proposed for self-gravitating systems against a static background, as detailed in \cite{Herrera:2018bww}. This paper introduces a modern interpretation of complexity, primarily motivated by addressing two deficiencies in earlier descriptions:
i) The prior descriptions lacked a comprehensive inclusion of all components of the energy density fluid, focusing solely on the energy density itself while ignoring potential additional contributions such as pressure.
ii) These earlier models replaced the probability distribution with the energy density of the fluid distribution, as was shown in \cite{Sanudo:2008bu}.

The implementation of the complexity factor formalism can be divided into two parts, the first, devoted to a purely mathematical interest (as can be seen in
\cite{Sharif:2018pgq,Sharif:2018efi,Abbas:2018cha,Herrera:2019cbx} and references therein), and the second one, where the idea is used to correct a weakness present in the study of stellar interiors, i.e. the necessity to close the system via the inclusion, usually by hand, of some well-behaved mass/density profile (see for instance \cite{Panotopoulos:2018joc,Panotopoulos:2018ipq,Moraes:2021lhh,Gabbanelli:2018bhs,Panotopoulos:2019wsy,Lopes:2019psm,Panotopoulos:2019zxv,Abellan:2020jjl,Panotopoulos:2020zqa,Bhar:2020ukr,Panotopoulos:2020kgl,Panotopoulos:2021obe,Panotopoulos:2021dtu} and references therein). 
In this sense, it is obvious that the second option is a strong motivation to investigate how the complexity factor formalism can also complete the set of equations in the context of stellar interiors.
In other words, the vanishing complexity factor formalism (i.e. a simplified version of the complexity factor formalism) can be used as a tool that allows us to complete the system and thus obtain novel spherically symmetric solutions.
Thus, the vanishing complexity formalism provides an alternative way of dealing with anisotropies in the context of compact stars (see for example \cite{Arias:2022qrm,Andrade:2021flq,Rincon:2023zlp,Rincon:2023ens}).

In what follows, we will only summarise the minimal information needed to make the article self-contained. Please note that we have only highlighted a few steps and that the detailed derivation must be consulted in \cite{Herrera:2018bww}.

Let us first note that the basic idea of this approach, at least when applied directly to the context of relativistic compact stars, is to approach the problem of anisotropy from an alternative perspective. 
It is well known that many relevant attempts have been made to explore the properties of realistic compact stars, describing either static, stationary or collapsing relativistic compact objects. In general, most of these approaches are coordinate-dependent. However, the complexity factor formalism (based on the orthogonal decomposition of the Reimann tensor) provides coordinate-free results expressed in terms of structure scalars, which are closely related to the kinematic and physical properties of the fluid \cite{Ospino:2020loc}.
Even more, we should recognize that for static anisotropic matter configurations, the complexity factor approach is a geometric mechanism for obtaining equations of state that link the radial and tangential pressures \cite{Ospino:2020loc}.
Technically speaking, the main idea is the orthogonal decomposition of the Riemann tensor for static self-gravitating fluids with spherical symmetry (see also \cite{Gomez-Lobo:2007mbg}). 
The decomposition considers three fundamental tensors:
i) $Y_{\alpha \beta}$
ii) $Z_{\alpha \beta}$
and
iii) $X_{\alpha \beta}$.
defined in terms of the Riemann tensor by
\begin{eqnarray}
Y_{\alpha \beta} &=& R_{\alpha \gamma \beta \delta}u^{\gamma}u^{\delta}, \label{electric} 
\\    
Z_{\alpha \beta} &=& R^{*}_{\alpha \gamma \beta
\delta}u^{\gamma}u^{\delta} = \frac{1}{2}\eta_{\alpha \gamma
\epsilon \mu} R^{\epsilon \mu}_{\quad \beta \delta} u^{\gamma}
u^{\delta}, \label{magnetic} 
\\
X_{\alpha \beta} &=& R^{*}_{\alpha \gamma \beta \delta}u^{\gamma}u^{\delta}=
\frac{1}{2}\eta_{\alpha \gamma}^{\quad \epsilon \mu} R^{*}_{\epsilon
\mu \beta \delta} u^{\gamma}
u^{\delta}. \label{magneticbis}
\end{eqnarray}
Notice that symbol $*$ is the dual tensor, in other words, 
\begin{equation}
   R^{*}_{\alpha \beta \gamma \delta}=\frac{1}{2}\eta_{\epsilon \mu \gamma \delta}R_{\alpha \beta}^{\quad \epsilon \mu} ,
\end{equation}
and $\eta_{\epsilon \mu \gamma \delta}$ has the usual meaning, i.e., it is the Levi-Civita tensor.
Now, the next step is to connect the tensors $Y_{\alpha \beta}, Z_{\alpha \beta}, X_{\alpha \beta}$ with the physical variables, i.e., 
\begin{eqnarray}
Y_{\alpha\beta} &=& \frac{4\pi}{3}(\rho +3
P)h_{\alpha\beta}+4\pi \Pi_{\alpha\beta}+E_{\alpha\beta},\label{Y}
\\
Z_{\alpha\beta} &=& 0,\label{Z}
\\
X_{\alpha\beta} &=& \frac{8\pi}{3} \rho
h_{\alpha\beta}+4\pi
 \Pi_{\alpha\beta}-E_{\alpha\beta}.\label{X}
\end{eqnarray}
At this point, it is essential to point out two auxiliary quantities, $\{ E_{\alpha \beta} \equiv C_{\alpha \gamma \beta \delta}u^{\gamma}u^{\delta} , E \}$ which are defined according to:
\begin{align}
E_{\alpha \beta} &= E \bigg(s_\alpha s_\beta+\frac{1}{3}h_{\alpha \beta}\bigg),
\label{52bisx}
\\
E &= -\frac{e^{-\lambda}}{4}\left[ \nu ^{\prime \prime} + \frac{{\nu
^{\prime}}^2-\lambda ^{\prime} \nu ^{\prime}}{2} -  \frac{\nu
^{\prime}-\lambda ^{\prime}}{r}+\frac{2(1-e^{\lambda})}{r^2}\right].
\label{defE}
\end{align}
The tensor $E_{\alpha \beta}$ must satisfy the following properties:
 \begin{eqnarray}
 E^\alpha_{\,\,\alpha}=0,
 \quad 
 E_{\alpha\gamma} =  E_{(\alpha\gamma)},
 \quad 
 E_{\alpha\gamma}u^\gamma=0.
  \label{propE}
 \end{eqnarray} 
Then, in order to show how the complexity factor arises, it is essential to realize that the set $\{ Y_{\alpha\beta}, Z_{\alpha\beta}, X_{\alpha\beta} \}$ can be parameterized by auxiliary scalar functions, as already pointed out in \cite{Herrera:2009zp}. Therefore, in the following, we will limit our interest to two tensors: $X_{\alpha \beta}$ and $Y_{\alpha \beta}$ (in the static case). Such tensors can be decomposed into four scalars, the so-called structure scalars $X_T, X_{TF}, Y_T, Y_{TF}$. Of course, it is possible to relate them to the physical variables of the system, as follows:
\begin{align}
X_T     &= 8\pi  \rho ,  \label{esnIII} 
\\
X_{TF}  &= \frac{4\pi}{r^3} \int^r_0{\tilde r^3 \rho ' d\tilde r} \label{defXTFbis}, 
\\
Y_T     &= 4\pi( \rho  + 3 P_r-2\Pi) \label{esnV}, 
\\
Y_{TF}  &= 8\pi \Pi- \frac{4\pi}{r^3} \int^r_0{\tilde r^3 \rho' d\tilde r} \label{defYTFbis}.
\end{align}
Having reached this point, it should be noted that the anisotropy factor can be written in terms of two of the previously introduced scalars, i.e.
\begin{equation}
8\pi \Pi = X_{TF} + Y_{TF} \label{defanisxy}.
\end{equation}
Last but not least, the scalar $Y_{TF}$ is then named as complexity factor and when it acquires a null value a relation between the energy density and the anisotropic factor is immediately obtained. Thus, demanding $Y_{TF}=0$ we found
\begin{equation}
\Pi(r) = \frac{1}{2r^3} \: \int^r_0{\tilde r^3 \rho'(\tilde r) d\tilde r} .
\end{equation}
It is worth mentioning that the approach based on the vanishing complexity factor formalism implies a certain relation between the energy density profile and the anisotropic factor, which is always the same for all matter content, and so it is universal and it does not depend on the details of the microphysics, such as particle species, nature of interactions etc. This is not surprising since it is a geometrical approach, and it must be contrasted to the more conventional approaches, where it is expected that the form of the anisotropy will be different from one type of stars to another, depending on the matter content, spin-statistics, form of the interaction potential and so on and so forth.

Notably, the vanishing complexity factor condition $Y_{TF} = 0$ is considered as a special case of the dynamic adiabatic scenario in compact stars $Z_{\alpha \beta} = 0$, and it plays a relevant role because the complexity factor takes into account the influences of energy density inhomogeneities and local anisotropy of pressures on the active gravitational (Tolman) mass (see \cite{Ospino:2020loc}). Thus, the vanishing complexity factor condition assumes $Z_{\alpha \beta} = 0$.
It should be pointed out that 
i) $X_{T}$ defines the energy density,
ii) $X_{TF}$ controls the inhomogeneity of the energy density (in the absence of dissipation),
iii) $Y_{TF}$ encodes the influence of both energy density 
inhomogeneity and local anisotropy of the pressure on the Tolman mass, and finally
iv) $Y_{T}$ is identified as the Tolman mass density (see \cite{Herrera:2009zp}
for further details).

Before concluding this part it is suitable to highlight a few points:
i) the anisotropic factor for a self-gravitating system can be directly computed after providing a density profile $\rho(r)$.
ii) the connection between $\Pi(r)$ and $\rho'(r)$ allows us to obtain an integro-differential equation that can be directly solved using numerical methods. Thus, such an equation becomes:
\begin{align}
    P'_r=-\frac{(m + 4 \pi P_r r^3)}{r \left(r - 2m\right)}\left( \rho + P_r\right)+
    \frac{1}{r^4} \int^r_0{\tilde r^3 \rho'(\tilde r) d\tilde r}.
\end{align}
Finally, given the notorious advantage of considering the (vanishing) complexity factor formalism in the context of compact stars, in this paper, we will investigate how the implementation of this approach can modify parameters that are useful to get insights about the star, the Tidal Love Numbers. In the next section, we will introduce them summarizing the required expression as well as the relevant results.

\section{Tidal Love Numbers}\label{TLN}

In the following, we will give the relevant expressions and the basic comments on the tidal love numbers. 
Let us start with a system of two individual bodies (let us introduce them, for example, as a primary and a secondary body) that are affected by the gravitational field produced by each other.
The gravitational potential of the secondary body (at large distances) can be obtained directly and written in terms of two contributions: 
 i) the conventional Newtonian potential term, and
ii) corrections given by the "influences are of third order in the ratio of the radius of the configuration to the distance between the components" \cite{1933MNRAS..93..449C}.
Thus, we have a result different from the trivial (Newtonian case) when the secondary object is distorted by the tidal field due to the primary one, and this precisely occurs when the principal moments of inertia of the secondary at its centre of gravity $\{A', B', C'\}$ differ from the moment of inertia about the radius vector $I'$, namely, $(A'-I'), (B'-I'), (C'-I')$ diff $0$.
In other words, the interaction of these two bodies implies the existence of tides on the surface of stars, in both the relativistic and the non-relativistic case.

A complete description of the theory of tidal Love numbers can be consulted in \cite{Hinderer:2007mb,Damour:2009vw,Postnikov:2010yn}.
The tidal Love number $k$, a quadrupole moment number and dimensionless coefficient, depends on the structure of the star and thus on its mass and equation of state. It is directly related to two auxiliary quantities commonly referred to as "deformabilities", denoted $\tilde{\lambda}$ and $\Lambda$, which are defined as follows:
\begin{align}
\tilde{\lambda} &\equiv \frac{2}{3} k R^5,
\label{eq:Love1}
\\
\Lambda &\equiv \frac{2 k}{3 C^5},
\label{eq:Love2}
\end{align}
where $C=M/R$ is the well-known compactness factor of the star.
Subsequently, the tidal Love number can be written in terms of $C$ as follows given by \cite{Hinderer:2007mb,Damour:2009vw,Postnikov:2010yn}
\begin{align}
k &= \frac{8C^5}{5} \: \frac{K_{o}}{3  \:K_{o} \: \ln(1-2C) + P_5(C)} ,
\label{elove}
\\
K_{o} &= (1-2C)^2 \: [2 C (y_R-1)-y_R+2] ,
\\
y_R &\equiv y(r=R) ,
\end{align}
where $P_5(C)$ is a fifth-order polynomial given by
\begin{align}
\begin{split}
P_5(C) = \: & 2 C \: \Bigl( 4 C^4 (y_R+1) + 2 C^3 (3 y_R-2) \ + 
\\
&
2 C^2 (13-11 y_R) + 3 C (5 y_R-8) -
\\
&
3 y_R + 6 \Bigl)  ,
\end{split}
\end{align}
%
%
and the function $y(r)$ is solution of a Riccati differential equation \cite{Postnikov:2010yn}:
\begin{align}
\begin{split}
r y'(r) + y(r)^2 &+ y(r) e^{\lambda (r)} \Bigl[1 + 4 \pi r^2 ( p(r) - \rho (r) ) \Bigl] 
\\
&+ r^2 Q(r) = 0 ,
\end{split}
\end{align}
supplemented by the initial condition (at the centre), $y(0)=2$, where
\begin{align}
\begin{split}
  \displaystyle Q(r) = 4 \pi e^{\lambda (r)} \Bigg[ 
  5 \rho (r) 
  &+ 9 p(r) + \frac{\rho (r) + p(r)}{c^2_s(r)} 
  \Bigg] 
  \\
  &- 6 \frac{e^{\lambda (r)}}{r^2} - \Bigl[\nu' (r)\Bigl]^2 .
\end{split}
\end{align}
Notice that $c_s^2 \equiv dp/d\rho = p'(r)/\rho'(r)$ is the speed of sound. 

It has been pointed out in \cite{Damour:2009vw,Postnikov:2010yn,Hinderer:2009ca} that when phase transitions or density discontinuities are present, it is necessary to slightly modify the expressions above. Since in the present work the energy density takes a non-vanishing surface value, in our analysis we incorporated the following correction
\begin{equation}
y_R \rightarrow y_R - 3 \frac{\Delta \rho}{\langle \rho \rangle},
\end{equation}
where $\Delta \rho$ is the density discontinuity, and 
\begin{equation}
\langle \rho \rangle = \frac{3M}{4 \pi R^3},
\end{equation}
is the mean energy density throughout the object. 

It is easy to see that since $k \propto (1-2C)^2$, tidal Love numbers of black holes vanish due to the fact that the factor of compactness of black holes $C=1/2$. This is an intriguing result of classical GR saying that tidal Love numbers of black holes, as opposed to other types of compact objects, are precisely zero. Therefore, a measurement of a non-vanishing $k$ will be a smoking-gun deviation from the standard black hole of GR. Moreover, regarding gravity  waves observatories, the Einstein Telescope will pin down very precisely the EoS of neutron stars \cite{Iacovelli:2023nbv}, while the Laser Interferometer Space Antenna is able to probe even extremely compact objects (with a factor of compactness $C > 1/3$), and it will set constraints on the Love numbers of highly-spinning central objects at $\sim 0.001-0.01$ level \cite{Piovano:2022ojl}.

\begin{table*}
\caption{Dimensionless deformability, $\Lambda$, for several stellar masses i) within the conventional approach for three different values of $\kappa$, i.e.: $\kappa=-0.45$, $\kappa=-0.90$, $\kappa=-1.35$ and ii) within the vanishing complexity factor formalism.}
\label{tab:1}
\begin{tabular*}{\textwidth}{@{\extracolsep{\fill}}ccccc@{}} 
\hline
Stellar mass ($M_{\odot}$) & $\kappa=-0.45$ & $\kappa=-0.90$ & $\kappa=-1.35$ &  Complexity  \\ 
\hline
1.0 & 3029.430 & 2996.260 & 3000.150 &  2744.770  \\ 
1.1 & 1991.840 & 1993.000 & 1989.510 &  1795.110  \\ 
1.2 & 1363.290 & 1358.290 & 1354.110 &  1201.550   \\ 
1.3 & 951.575  & 945.440  & 941.301  &  818.306   \\ 
1.4 & 675.674  & 669.815  & 665.975  &  563.446   \\ 
1.5 & 486.864  & 481.504  & 477.706  &  388.963    \\ \hline
\end{tabular*}
\end{table*}

\section{Results and Discussion}\label{Dis}

In the present paper we have investigated within GR anisotropic stars made of dark energy in light of the vanishing complexity formalism. In particular, we compute numerically interior solutions of realistic spherical configurations of anisotropic matter, and we compare our solution against a more conventional approach, i.e. assuming an ansatz for the anisotropic factor by hand. Both the TOV equations and the Riccati differential equation have been integrated numerically with the help of the well-known and widely used Wolfram Mathematica. We have developed simple codes that do the numerical work in a straightforward manner, and they do not require a lot of time. As a matter of fact, for each variation the code runs, computes and generates the plots within $\sim 5 \: sec$.

For dark energy, following \cite{DEstars2}, we consider the extended Chaplygin equation of state \cite{Pourhassan:2014ika,Pourhassan:2014cea}.
\begin{equation}
    \displaystyle p(r) = - \frac{B^2}{\rho (r)} + A^2 \rho (r) ,
\end{equation}

\medskip

\noindent where $A$ and $B$ are positive constants, out of which $A$ is dimensionless, while $B$ has dimensions of energy density and pressure. Also, note that Chaplygin's EoS is given by $p = -B^2/\rho$ \cite{chaplygin1}, while the additional barotropic term corresponds to $A^2\rho$ leading also to a good dark energy model \cite{Panotopoulos:2020qbx}. The numerical values of $A$ and $B$ considered here are as follows
\begin{equation}
A = \sqrt{0.45}, \; \; \; \; B = 0.2 \times 10^{-3}~km^{-2}.
\end{equation}

\medskip

It must be taken into account that given the form of the EoS, the pressure and density cannot become zero at the same time. Therefore, a vanishing pressure at the surface of the star implies a non-vanishing value for the energy density, which is found to be $\rho_s = B/A$.

\smallskip

As a supplementary check, we have obtained, numerically again, interior solutions using a more standard approach, i.e., adding external constraints to close the system of differential equations. As a toy model, we have considered an anisotropic factor, $\Pi(r)$, as follows \cite{Silva:2014fca,Folomeev:2015aua,Cattoen:2005he,Horvat:2010xf,Arbanil:2021ahh}
\begin{equation} \label{standard}
\Pi(r) = \kappa \: p_r(r) \: \Bigl( 1-e^{-\lambda(r)} \Bigl)
\end{equation}
characterized by a dimensionless parameter, $\kappa$, which encodes the strength of the anisotropy. Also, we consider three different values, $\kappa=-0.35, -0.9, -1.35$.

\smallskip
Our numerical calculations, summarized in two figures, reveal the following facts:
\begin{itemize}
\item[$-$]
i) In the mass-radius profile $M(M_{\odot})$ vs $R(km)$, the vanishing complexity factor condition $(Y_{TF}=0)$ implies that we have a more compact star, at least with respect to the anisotropic star we used for comparison. In other words, the maximum mass $M_{\text{max}}$ and the maximum radius $R_{\text{max}}$ take smaller values for each type of matter. Additionally, in scenarios involving the conventional anisotropy equation, specifically when employing Eq.\eqref{standard}, it is observed that a more negative value of the dimensionless parameter $\kappa$ yields results that increasingly align with those derived under the vanishing complexity factor condition.
\item[$-$]

ii) We conclude that tidal deformability is pivotal in observing coalescing stars through gravitational waves, serving as a key tool for unveiling the inner structure of specific types of relativistic stars, including neutron stars. Since tidal Love numbers characterize how easy (or difficult) it would be for a star to deform away from sphericity, we find that $k$ gradually decreases with increasing $C$ up to a certain value, within the vanishing complexity factor formalism and also adopting a more standard approach assuming an anisotropic factor given by Eq. \eqref{standard}. Thus, we found that although tidal Love numbers are quite similar for small values of $C$, beyond a certain threshold, the vanishing complexity factor conditions do not produce lower values of $k$, which means that anisotropic stars in this framework retain a reminiscent deformation, since $k$ of compact stars never vanishes (the limit $k \rightarrow 0$ as $C \rightarrow 1/2$ corresponds to black holes).
\end{itemize}
Thus, in light of those numerical results, we conclude that the complexity factor formalism is a powerful tool for inferring novel features in the context of compact stars.


\begin{figure*}[ht!]
\centering
\includegraphics[width=0.75\textwidth]{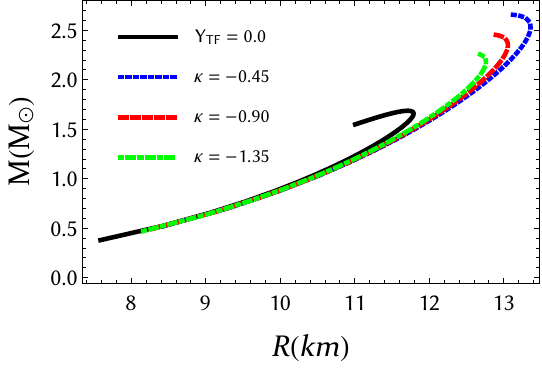} \
\caption{
Mass-to-radius relationships for the anisotropic stars investigated in this work. We have considered i) vanishing complexity factor formalism (black) and ii) a conventional approach for 3 different values of the parameter $\kappa=-0.45,-0.90,-1.35$ from top to bottom, see text for more details.
}
\label{fig:1} 	
\end{figure*}


\begin{figure*}[ht!]
\centering
\includegraphics[width=0.75\textwidth]{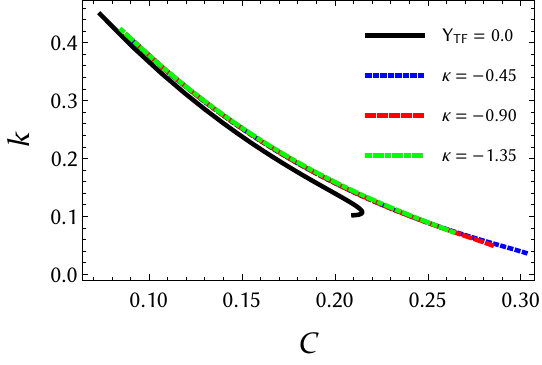} 
\caption{
Tidal Love number $k$ versus factor of compactness for the anisotropic stars investigated in this work. We have considered i) the vanishing complexity factor formalism (black), and ii) a more conventional approach for 3 different values of the parameter $\kappa=-0.45,-0.90,-1.35$, see text for more details.
}
\label{fig:2} 	
\end{figure*}



\section{Conclusions}\label{Conc}

\smallskip\noindent

In this study, we have computed the quadrupolar gravitoelectric tidal Love numbers for spherical configurations composed of anisotropic matter within the framework of General Relativity. Anisotropies were introduced via the vanishing complexity factor formalism \cite{Herrera:2018bww}, and interior solutions were obtained using the Extended Chaplygin gas equation of state
 \cite{2002hep.th....5140B,Pourhassan:2014ika}. 
 We compared our approach with a more conventional method where anisotropy is introduced through a prescribed anisotropic factor \cite{Silva:2014fca,Folomeev:2015aua}.

\smallskip\noindent	

Our numerical results indicate that the vanishing complexity factor condition leads to more compact stellar configurations compared to the conventional anisotropy models. Specifically, both the maximum mass and radius of the stars are smaller under the vanishing complexity condition. Additionally, we observed that the tidal Love numbers decrease with increasing compactness up to a certain threshold. Beyond this point, the vanishing complexity factor condition does not yield lower tidal Love numbers, suggesting that anisotropic stars within this framework retain a residual deformation. This contrasts with black holes, for which the tidal Love numbers vanish
 \cite{2009PhRvD..80h4018B,2019PhRvD..99b4036C,2016CQGra..33i5005Y}, 
 and implies that such anisotropic stars could potentially be distinguished from black holes through gravitational wave observations.
	
\smallskip\noindent	
These findings highlight the effectiveness of the vanishing complexity factor formalism in modelling anisotropic stars and emphasise the importance of anisotropy in determining the properties of compact objects. The residual tidal deformation suggests that gravitational wave observations, such as those from neutron star mergers detected by LIGO and Virgo \cite{2017PhRvL.119p1101A,2018PhRvL.121p1101A},
 could provide valuable insights into the anisotropic nature of dense matter.
	
\smallskip\noindent	

Future research could extend this work by exploring other equations of state, including those relevant for neutron stars and strange quark stars \cite{2005PrPNP..54..193W,2007PhR...442..109L,2023EPJC...83.1065R,2024JCAP...05..130R,2019ApJ...871..157S}, 
and by considering the effects of rotation and magnetic fields on the structure and tidal properties of anisotropic stars \cite{2013PhRvD..88b3009Y,2013PhRvD..88d4052D,2023ApJ...943...52R}. Moreover, investigating the stability and oscillation modes of these stars \cite{1999LRR.....2....2K,2020PhRvD.101f3025S,2023PhRvD.107l3022R} 
could provide further understanding of the impact of anisotropy on the dynamical behaviour of compact objects. Finally, incorporating observational data from current and future gravitational wave detectors could help constrain the models and improve our understanding of the role of anisotropy in astrophysical phenomena.

\section*{Acknowledgments}

The authors wish to thank the reviewer for useful comments and suggestions. A.~R. acknowledges financial support from 
Conselleria d'Educació, Cultura, Universitats i Ocupació de la Generalitat Valenciana thorugh Prometeo Project CIPROM/2022/13.
A.~R. is funded by the María Zambrano contract ZAMBRANO 21-25 (Spain) (with funding from NextGenerationEU).
I.~L. thanks the Funda\c c\~ao para a Ci\^encia e Tecnologia (FCT), Portugal, for the financial support to the Center for Astrophysics and Gravitation (CENTRA/IST/ULisboa)  through the Grant Project No.~UIDB/00099/2020  and Grant No.~PTDC/FIS-AST/28920/2017. 



%

\bibliographystyle{unsrt}
\bibliography{biblioIL.bib}

%

\end{document}